\documentstyle[12pt,aaspp4]{article}
\newcounter{firstfn}
\newcounter{secondfn}
\newcounter{thirdfn}
\newcounter{fourthfn}

\newcommand{\ifpp}[1]{#1}
\newcommand{\ifms}[1]{}
\ifpp{\input psfig}
\newcommand{\degree}{\ifmmode {^{\circ}} \else {$^{\circ}$} \fi}
\newcommand{\degrees}{\ifmmode {^{\circ}} \else {$^{\circ}$} \fi}

\newcommand{\unit}[1]{\ifmmode {\rm\ #1\,} \else {$\rm #1$} \fi}
\newcommand{\quarter}{\ifmmode {\frac{1}{4}} \else {$\frac{1}{4}$} \fi}
\newcommand{\angstrom}{\unit{\AA}}
\newcommand{\angstroms}{\unit{\AA}}

\newcommand{\etal}{{\em et al.~}}
\newcommand{\emunits}{\unit{{\rm cm}^{-6} {\rm pc}}}

\newcommand{\pressure}{\unit{{\rm cm}^{-3} {\rm K}}}

\newcommand{\K}{\unit{K}}

\newcommand{\lt}{\unit{<}}
\newcommand{\gt}{\unit{>}}

\newcommand{\tten}[1]{\ifmmode {\times 10^{#1}} \else {$\times 10^{#1}$} \fi}
\newcommand{\tentothe}[1]{\ifmmode {10^{#1}} \else {$10^{#1}$} \fi}
\newcommand{\parsec}{\unit{pc}}
\newcommand{\kpc}{\unit{kpc}}
\newcommand{\microgauss}{\unit{\mu G}}
\newcommand{\deriv}[2]{\frac{d#1}{d#2}}

\newcommand{\del}{\nabla}
\newcommand{\pu}{\unit{ph\ s^{-1}}\unit{cm^{-2}\ str^{-1}}}
\newcommand{\intensity}{\unit{erg\ s^{-1}\ cm^{-2}\ str^{-1}}}
\newcommand{\cu}{\unit{ph\ s^{-1}}\-\unit{cm^{-2}}\- \unit{str^{-1}\- \angstrom^{-1}}}
\newcommand{\HI}{{\rm H{\sc i\,}}}
\newcommand{\OI}{{\rm O{\sc i\,}}}
\newcommand{\ArI}{{\rm Ar{\sc i\,}}}
\newcommand{\CII}{{\rm C{\sc ii\,}}}
\newcommand{\SiII}{{\rm Si{\sc ii\,}}}
\newcommand{\MgII}{{\rm Mg{\sc ii\,}}}
\newcommand{\CIII}{{\rm C{\sc iii\,}}}
\newcommand{\NIII}{{\rm N{\sc iii\,}}}
\newcommand{\SiIII}{{\rm Si{\sc iii\,}}}
\newcommand{\OIII}{{\rm O{\sc iii}]\,}}
\newcommand{\CIV}{{\rm C{\sc iv\,}}}
\newcommand{\SIV}{{\rm S{\sc iv\,}}}
\newcommand{\SiIV}{{\rm Si{\sc iv\,}}}

\newcommand{\iovi}{{\rm \unit{I(OVI)}}}
\newcommand{\iciv}{{\rm \unit{I(CIV)}}}

\newcommand{\OVI}{{\rm O{\sc vi\ }}}
\newcommand{\NeVI}{{\rm Ne{\sc vi\,}}}
\newcommand{\ArVI}{{\rm Ar{\sc vi\,}}}
\newcommand{\doublet}{\ifmmode {\lambda\lambda} \else {$\lambda\lambda$} \fi}
\newcommand{\singlet}{\ifmmode {\lambda} \else {$\lambda$} \fi}
\newcommand{\cmsquared}{\unit{cm^2}}

\newcommand{\percmcubed}{\unit{cm^{-3}}}

\newcommand{\percmsqr}{\unit{cm^{-2}}}

\newcommand{\crateflux}{\unit{s^{-1} cm^{-2}}}

\begin{document}

\lefthead{Korpela, Bowyer, \& Edelstein}
\righthead{Diffuse FUV Emission from the ISM}

\title{Spectral Observations of Diffuse FUV Emission from the Hot Phase
of the Interstellar Medium with DUVE (the Diffuse Ultraviolet Experiment)}
\author{Eric J. Korpela, Stuart Bowyer, and Jerry Edelstein}
\affil{Space Sciences Laboratory, University of California, Berkeley, CA 94720}

\begin{abstract}
One of the keys to interpreting the character and evolution of
interstellar matter in the galaxy is understanding the distribution
of the low density hot ($10^5 \K -10^6 \K$) phase of the interstellar medium (ISM).
This phase is much more difficult to observe than the cooler high density
components of the ISM
because of its low density and lack of easily observable tracers.
Because gas of this temperature emits mainly in the 
far ultraviolet (912 \angstrom - 1800 \angstroms) and extreme ultraviolet
(80 \angstrom - 912 \angstrom), and (for gas hotter than $10^6$ K) X-rays, 
observations 
in these bands can provide 
important constraints to the distribution of this gas.  Because of 
interstellar opacity at EUV wavelengths, only FUV and X-ray observations can provide clues to the
properties of hot gas from distant regions.
We present results from a search for FUV emission from the diffuse ISM
conducted with an orbital FUV spectrometer, DUVE, which was launched in
July, 1992.  The DUVE spectrometer, which covers the band from 950 \angstrom
to 1080 \angstrom with 3.2 \angstrom resolution, observed a region of low
neutral hydrogen column density near the south galactic pole for a total
effective integration time of 1583 seconds.  The only emission line detected
was a geocoronal hydrogen line at 1025 \angstrom.  We are able to
place upper limits to several expected emission features that provide constraints
on interstellar plasma parameters.  We are also able to place limits on the
continuum emission throughout the bandpass.
We compare these limits and other diffuse observations with several models
of the structure of the interstellar medium and discuss the ramifications.
\end{abstract}

\keywords{ISM: general, ultraviolet: ISM, instrumentation: spectrographs, space vehicles}

\section{Introduction}

Since the prediction of the existence of hot gas in the interstellar
medium (ISM) by Spitzer (1956) and its subsequent detection
by Bowyer, Field and Mack (1968) in Soft X-ray emission, a variety
of models have been developed which attempt to explain the source of this
gas, and how it evolves over time.  
The three dominant types of model, 
Galactic fountain models (\cite{shapiro76}), three phase models (\cite{mckee77}), and 
isolated supernova type models (\cite{smith74}), all claim some measure of success in fitting the
existing observations.

In the past several decades, measurements of interstellar absorption in
the far ultraviolet (FUV) spectra of many stars 
using Copernicus and IUE (Jenkins 1978a\&b, Savage \& Massa 1987) 
have resulted in the detection of many highly
ionized species.
 More recently
the Goddard High Resolution Spectrometer (GHRS) on the Hubble Space Telescope,
and the Berkeley FUV/EUV Spectrometer on ORFEUS  
(\cite{sembach95,spitzer96,hurwitz96}) have provided additional data in this
area. 

Absorption measurements are
fairly insensitive to the distribution of absorbing material along the line
of sight and do not provide enough information
to distinguish between various models of the ISM.
Emission from the ISM, though difficult to detect, has been the subject of
various investigations.
Martin and Bowyer (1990) investigated emission in the FUV 
between 1300 and 1800 \angstrom using the Space Shuttle-borne UVX
experiment.  They detected \CIV \doublet 1548,1551 and \OIII \doublet 1661,1666
emission along several lines of sight.  

Several attempts have been made to measure diffuse line emission between
900 and 1200 \angstroms.  
Voyager observations made by Holberg produced an upper limit of about
$10^4$ \pu for \OVI \doublet 1032,1038 (\cite{holberg}).
However, these Voyager observations have recently been called into question by
Edelstein, Bowyer and Lampton (1997).  
A similar limit for the \OVI doublet of $1.4\tten{4} \pu$ was also reached by 
Edelstein 
and Bowyer (1993).
More recently Dixon \etal (1996) used the Hopkins Ultraviolet Telescope
(HUT) to search for diffuse FUV emission.
They observed ten
lines of sight at high galactic latitude and claim detections of \OVI 
emission along four
of the ten lines of sight.
Dixon \etal determined that the \OVI\ they 
measured
was consistent with temperatures and pressures
which are as much as an order
of magnitude higher than those determined by Martin and Bowyer's \CIV 
measurements.  Therefore they concluded that the \OVI they saw was produced
by gas that is distinct from that which produces \CIV emission.

\section{The Instrument and Calibration}

We have designed an instrument capable of measuring the
important \OVI \doublet 1032,1038 emission from 
the diffuse interstellar medium.
The instrument's small bandpass (150
\angstrom) around the \OVI lines also includes the potentially important
\CIII \singlet 977, \NIII \singlet 991, and \CII \singlet 1037 lines.
Many problems need to be solved in the design of a diffuse spectrometer for
this bandpass.
The low predicted intensity of these emission lines requires
that the
spectrometer have a high effective area $\times$ solid angle
product.  In addition, sources of instrument background and spectral
contamination must be limited.  We discuss, in turn, important
sources of background and the techniques used to mitigate them.

There are several bright airglow lines 
which could interfere with attempts to observe the faint \OVI \doublet 
1032,1038 lines. Predictions of \OVI emission indicate
the intensity of these line should be near 5000 \pu (\cite{shull}).
The airglow lines nearest to the \OVI doublet, \HI \singlet 1025 and \OI \singlet 1027,
have a combined intensity of about $10^{5.5}$ \pu. Assuming Gaussian line profiles,
it can be shown that a spectral resolution of 2.7 \angstrom half-energy width 
(HEW) is required if the contribution
to a bin at 1032 \angstrom is to contain less than 5000 \pu equivalent 
background due to these lines.  For this reason, high spectral resolution
is required to prevent contamination of the faint interstellar emission lines.

An even more troubling background source is the far brighter 
\HI \singlet 1216 line.  With 
an intensity of 5 kiloRaleighs (4 \tten{8} \pu), scattering 
within the instrument
could easily swamp any signals. \HI \singlet 1216 photons have to be rejected
to a level of 5 \tten{-6} per spectral bin or less to have an equivalent
contribution of 5000 \pu.  Unfortunately no filter
material exists which is able to block \HI \singlet 1216 while allowing \OVI \doublet 1032,1038
to pass.  A double spectrometer method was used to limit \HI \singlet 1216 scattering
as described below.

Another potential contaminant in the diffuse FUV spectrum is stellar 
flux.
Bright O, B, and A stars can be eliminated simply by
not pointing the instrument
near any known bright
stars of this type.  
Emission from later FUV emitting stars is more difficult to deal with.  
We designed our instrument with imaging along the slit to allow identification
of stellar signals.
We also used an additional detector which provided an FUV image of
the spectrometer field.  This allowed detection of stars too faint
to be seen directly in the spectra.

A background due to exospheric charged particles is
potentially quite substantial.  This background can
range from 1 \crateflux  to upwards of \tentothe{5} \crateflux for an
open faced microchannel plate detector.  
It is highly dependent on orbital parameters,
especially ram angle, solar angle, altitude and position over the earth's
surface.  
In our instrument, the entrance aperture is covered with a fine mesh charged to +28 Volts 
to limit the number of exospheric charged particles that can reach the detector.
In 
addition, an opaque shutter 
was employed
to provide an in-flight determination of the entire non-photonic background.

The instrument, designated  DUVE, the Diffuse Ultraviolet Experiment, is 
based on a spectrometer previously developed by Edelstein and Bowyer (1993).
However, the DUVE  instrument's capability for studying diffuse radiation was substantially
improved.
The basic design is a two stage spectrometer which solves many
of the background problems described above, yet provides a large area $\times$ solid
angle product when compared to alternative designs.  
A schematic of the instrument is shown
in Figure~\ref{fig5}.
\ifpp{
\begin{figure}
\begin{center}
\ \psfig{file=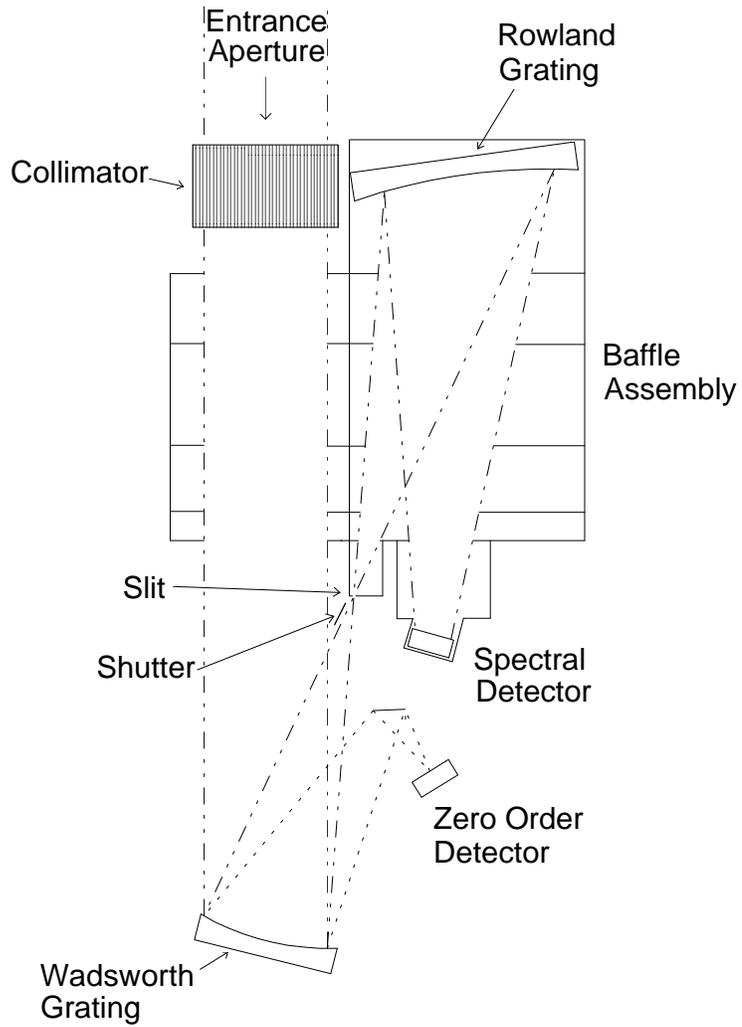,width=5.5in}
\caption{A schematic of the DUVE instrument. The dot-dash
lines represent the light path from entrance aperture to spectral detector.
The dotted line represents the path of zero order light from the Wadsworth
grating.\label{fig5}}
\end{center}
\end{figure}
}

The first stage of the spectrometer filters out unwanted
wavelengths, most notably \HI \singlet 1216.  Light entering the spectrometer
passes through a wire grid collimator with a full width of $\pm 1 \degree$.
After passing through the collimator, the light strikes a diffraction 
grating in Wadsworth configuration.  
First order light from the grating is focused toward a slit.
The slit width defines the horizontal (across the slit) sky acceptance angle.
Contrary to intuition, this slit does not limit the instrument bandwidth.
The collimator's angular width and the Wadsworth grating ruling density 
define the width of
the instrument bandpass. Each individual wavelength
accepted by the slit has entered the collimator at a slightly different angle.
This has the unique advantage of limiting contamination by a bright star
to a wavelength range much smaller than the overall bandpass.

The second stage of the spectrometer is a dispersion stage.  Light
entering through the slit strikes a holographically corrected diffraction 
grating in a 
Rowland
circle configuration.  The second order diffracted light from the grating
is focused onto a microchannel plate detector.
Use of second order light in the diffraction stage allows high dispersion
without significantly increasing the line density required to achieve the
required spectral resolution.
The spectral resolution of
the second stage is determined by properties of the diffraction grating
and the width of the entrance slit.
Unwanted orders are blocked in this stage of the spectrometer through use
of strategically placed baffles.  They are specifically designed to eliminate
all orders of \HI \singlet 1216 which may have been scattered through the 
entrance
slit.

Extensive calibration of the DUVE instrument was carried out using the
EUV/FUV calibration facilities at the Space Sciences Laboratory (\cite{ssl}).
The details of this calibration are beyond the scope of this paper, but are 
described in detail by Korpela (1997).  In brief, all the components needed
to calculate the effective (efficiency$\times$area$\times$solid angle) product
(the instrument grasp) were measured at a variety of wavelengths. The result
is shown in Figure~\ref{fig9}.

\ifpp{
\begin{figure}
\begin{center}
\ \psfig{file=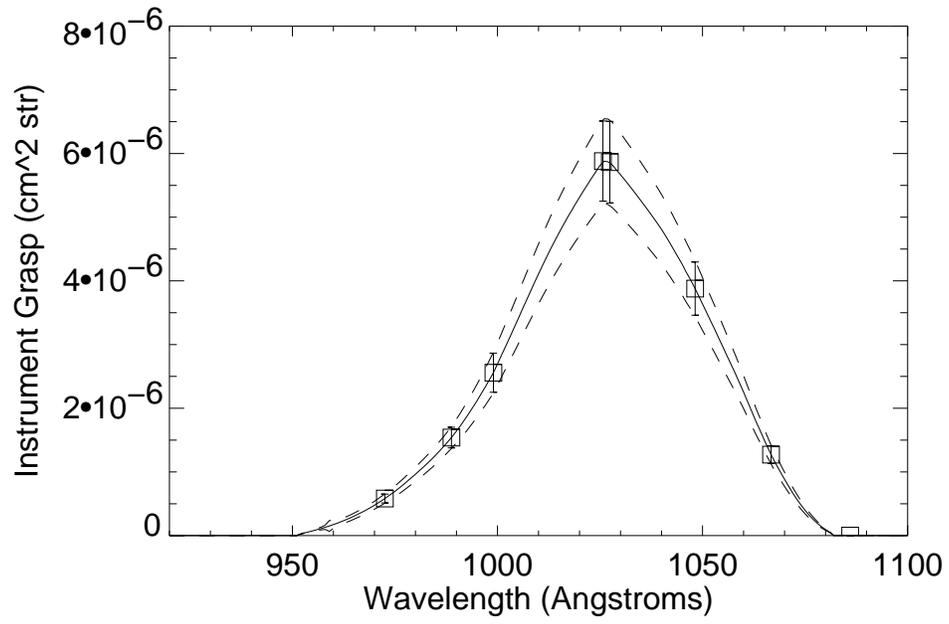}
\caption{DUVE area solid-angle product or grasp \label{fig9}}
\end{center}
\end{figure}
}

The spectral resolution was also measured.
The instrument was found to have a
HEW of 3.5 \angstroms at 1025 \angstroms.  However, the line profile was
found to be 
asymmetric with a much steeper cutoff toward long wavelengths. Consequently
the contribution of \HI \singlet 1025 into a spectral bin at 1032 \angstrom
is the same as an equivalent symmetric line with 
an HEW of 2.3 \angstroms.  The imaging focus along the slit was
also measured and was found to be $\sim 4$
arc minutes HEW.

\section{The Flight}

The DUVE instrument was launched as a secondary payload attached to the second
stage of a Delta II 7925 vehicle on 24 July, 1992.
As a secondary payload, 
the parameters of the DUVE orbit were determined by the requirements of
the primary payload, the Geotail satellite.  
The orbital apogee was 1460 km 
which placed the instrument well above the majority of atomic oxygen airglow
and above about half of the expected geocoronal hydrogen airglow.  

Due to safety requirements, the propellant and control gases of the second stage had to be depleted 
by the end of the first orbit.  Therefore the mission was conducted
in two phases, a pointed phase for the first orbit, and a spin stabilized phase thereafter.  Because
the DUVE data were telemetered through the second stage telemetry system,
the mission lifetime for DUVE was limited by the battery power available to the
telemetry system (about 8 hours total).

During the first orbital night, the instrument was pointed at the
primary target and a slow drift (0.1 \degree min$^{-1}$ perpendicular to the slit)
was initiated.  This phase continued until T+5000 seconds, at which time
second stage spin up was begun.  Observations continued
throughout this period, however, much of the data were contaminated
by stars and bright airglow.
The primary target was an area of low neutral hydrogen column density and
enhanced soft X-ray emission near $l=10^\circ$, $b=-60^\circ$.  Because of
the spin and its precession throughout the orbit, $\sim 2.3$ steradians of
sky was sampled during the observations. 

In
flight, overall background rates ranged from 1 \crateflux 
to 20 \crateflux with a median value of 
2 \crateflux, indicating that the methods employed to reduce the
background in the instrument
were quite successful.
Overall, our backgrounds consisted of roughly  equal parts charged
particle background and intrinsic detector background.  No scattered \HI \singlet 1216  
was detected.

\section{Observational Results}

Detector images (256 pixels spectral $\times$ 32 pixels imaging) were telemetered 
from the spacecraft during ground station contacts.
These images were telemetered multiple
times over many ground station passes, hence, many data dropouts were able to
be corrected.  The end result was 19 images obtained from 4151 seconds of
observation time.  These images were examined for stellar contamination,
uncorrected data dropouts and instrument anomalies and the affected
portions were masked off or otherwise corrected.  Where no
correction was possible, the images were discarded. 

The sky and shuttered background images are accumulated in an interleaved
fashion, each sky image has a corresponding background image.
The telemetered images were ordered in terms of 
count rate in the background image.
We then determined whether adding the next higher background image to the summed
low noise set would
decrease the overall noise rate ($\sqrt{\frac{N_{\rm bg}}{t}}$).  
Fifteen of the images
passed this test.  The background level of those images
ranged from 1.0
to 2.6 cm$^{-2}$ s$^{-1}$.  The background levels of those rejected ranged
from 8.2 to 20 cm$^{-2}$ s$^{-1}$.
\ifpp{
\begin{figure}
\begin{center}
\ \psfig{file=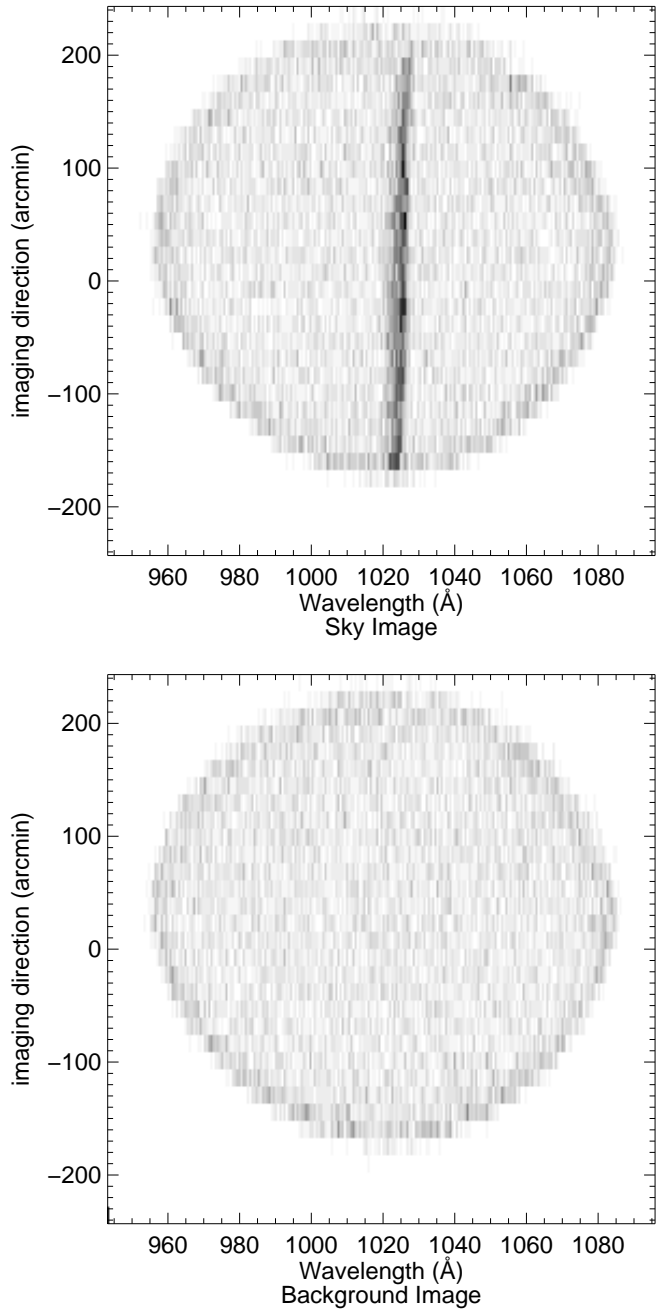}
\caption{Raw spectral detector images from the DUVE instrument's 
flight. The top image is a sum of shutter open images.  The vertical stripe
is \HI \singlet 1025 airglow.  The bottom image is a sum of the corresponding
shutter closed images.  
Integration times are 1583 and 1529 seconds respectively.
\label{fig11}} 
\end{center}
\end{figure}
}
The resulting sum of the accepted images is shown in Figure~\ref{fig11}.
The top image is a sum
of shutter opened images. The vertical stripe is \HI \singlet 1025 airglow.
The bottom is a sum of the corresponding shutter closed images. 
The image has not been corrected 
for distortions.

\ifpp{
\begin{figure}
\begin{center}
\ \psfig{file=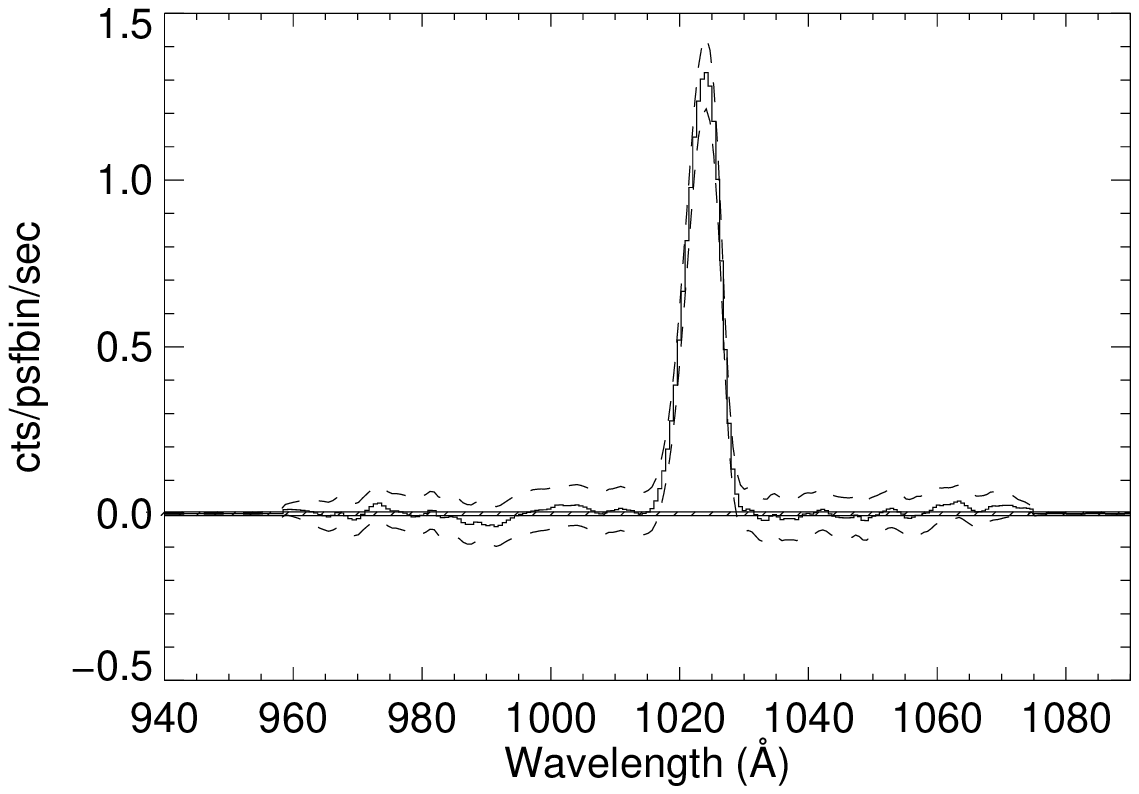}
\caption{The solid line represents
the flight spectral data after subtraction of the background image and 
convolution with a line spread function.
The dashed lines represent $\pm 3 \sigma$ error levels.\label{fig12}}
\end{center}
\end{figure}
}
The summed image was corrected for image distortions and shutter 
open/closed time, and the 
background image was subtracted from the data.  The resulting image
was histogrammed and convolved with a line spread function that was determined
during calibration.  
The convolved histogram is shown in Figure~\ref{fig12}.  The dashed lines
represent $\pm 3 \sigma $ error levels as determined through count statistics
and the error in the determination of the shutter open/closed time.  This
quantity is determined by:
\begin{eqnarray}
\sigma&=&\sqrt{\sigma_{\rm fg}^2+\sigma_{\rm bg}^2+\sigma^2_{\Delta t}}\nonumber \\
&=&{\sqrt{\left(\frac{N_{\rm fg}}{t_{\rm fg}}\right)^2
       +\left(\frac{N_{\rm bg}}{t_{\rm bg}}\right)^2
       +\left(\frac{N_{\rm fg}\Delta t}{t_{\rm fg}^2}
             +\frac{N_{\rm bg}\Delta t}{t_{\rm bg}^2}\right)^2}}
\end{eqnarray}
The quantities $N_{\rm fg}$ and $N_{\rm bg}$ are the number of counts (post-convolution) 
per wavelength bin in the sky and background images respectively.
The values $t_{\rm fg}$ and $t_{\rm bg}$ represent the shutter open and
shutter closed integration times (1582.6 and 1529.2 s)  and $\Delta t$
represents the uncertainty in those times (14.0 s).
The hatched area near the baseline in Figure~\ref{fig12}\ represents the
approximate fraction of the error that is due to potential errors in
shutter timing.

It can easily be seen that only one spectral line exceeds the $3 \sigma$ 
significance level.  That line is the geocoronal \HI \singlet 
1025 line.  By dividing the count rate in this line by the DUVE grasp,
we determined its intensity.  The 1.31 counts per second detected
corresponds to an intensity of $2.26\pm 0.26 \tten{5} \pu.$  As expected this
value is between that measured at low altitude (600 km) (\cite{chakrabarti}) 
and 
the interplanetary value (\cite{holberg}).
\ifpp{
\begin{figure}
\begin{center}
\ \psfig{file=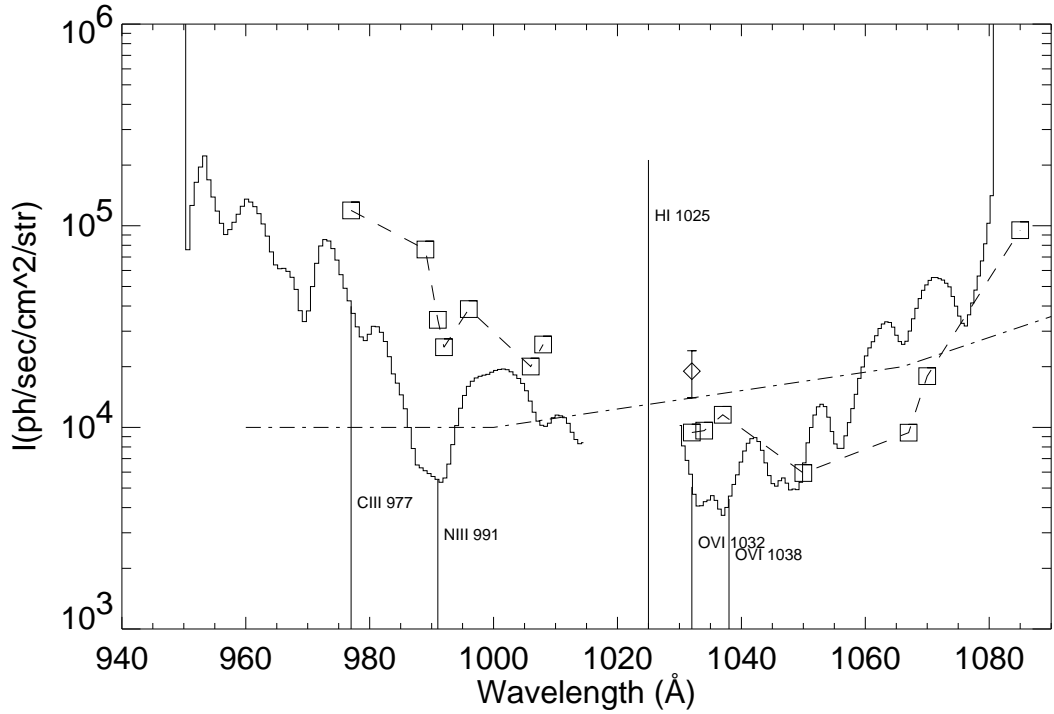}
\caption{Upper limits to line emission placed 
by this work are shown
as a solid line.  Previous limits by Edelstein and Bowyer (1993) are shown by
a dashed line. Limits determined from Voyager UVS measurements 
as analyzed by Edelstein, Bowyer, and Lampton (1997) are presented 
as a dot-dash line. The diamond with error bars
is a measurement of OVI 1032,1038 \angstrom emission by Dixon {\em et al.} (1996).  The vertical lines show the positions of several important astrophysical
lines.
\label{fig13}}
\end{center}
\end{figure}

\begin{table}
\begin{minipage}{5.95in}
\begin{center}
\begin{tabular}{crrr}
\hline
Species & \multicolumn{1}{c}{$\singlet (\angstrom)$}
        & \multicolumn{1}{c}{$I (\pu)$}
        & \multicolumn{1}{c}{$I (\intensity)$} \\  
\hline  
\hline  
\HI\footnotemark & 972 & $\leq$ 7.4\tten{4} & $\leq$ 1.5\tten{-6} \\
\setcounter{firstfn}{\value{footnote}}
\CIII & 977 & $\leq$ 4.0\tten{4} & $\leq$ 8.1\tten{-7} \\
\OI$^{\thefirstfn}$ & 989 & $\leq$ 6.1\tten{3} & $\leq$ 1.2\tten{-7} \\
\NIII & 991 & $\leq$ 5.5\tten{3} & $\leq$ 1.1\tten{-7} \\
\SiII & 992 & $\leq$ 5.7\tten{3} & $\leq$ 1.1\tten{-7} \\
\SiIII & 996 & $\leq$ 1.6\tten{4} & $\leq$ 3.2\tten{-7} \\ 
\NeVI & 1006 & $\leq$ 1.3\tten{4} & $\leq$ 2.6\tten{-7} \\
\ArVI & 1008 & $\leq$ 1.0\tten{4} & $\leq$ 2.0\tten{-7} \\
\HI$^{\thefirstfn,}$\footnotemark & 1025 & 2.26$\pm$0.26\tten{5} & 4.38$\pm$0.49
\tten{-6} \\ 
\setcounter{thirdfn}{\value{footnote}}
\OVI\footnotemark & 1032, 1038 & $\leq$ 7.6\tten{3} & $\leq$ 1
.4\tten{-7} \\
\setcounter{fourthfn}{\value{footnote}}
\CII& 1037 & $\leq$ 3.9\tten{3} & $\leq$ 7.4\tten{-8} \\
\ArI$^{\thefirstfn}$ & 1050 & $\leq$ 6.4\tten{3} & $\leq$ 1.2\tten{-7} \\ 
\SiIV & 1067 & $\leq$ 2.9\tten{4} & $\leq$  5.4\tten{-7} \\
\SIV & 1070 & $\leq$ 5.1\tten{4} & $\leq$ 9.5\tten{-7} \\  
\hline 
\end{tabular}
\caption{Upper limits to line emission placed by the DUVE data}
\label{tab4}
\end{center}
\footnotetext{$^\thefirstfn$ Anticipated airglow line}
\footnotetext{$^\thethirdfn$ Detected at 37$\sigma$.} 
\footnotetext{$^\thefourthfn$ This limit is total emission from the doublet
based upon joint statistics
by assuming $\frac{I(1032)}{I(1038)}=2$.  Upper limits for the individual
components of the doublet are $I(1032) \leq$ 5400 \pu and $I(1038) \leq$
4400 \pu.} 
\end{minipage}
\end{table}
}
We have used the DUVE data
to derive upper limits to line emission in the band covered by the spectrometer.
These limits are shown in Table~\ref{tab4}.
As a test of the error levels, fluctuations in the spectrum
outside of the 1025 \angstrom line were compared with those expected by
statistics.  A histogram of the standard deviation per wavelength bin, $\frac{\Delta N}{\sigma}$, when compared to the zero level was best
fit by a Gaussian of width $1.01\pm 0.03$ centered at $0.1\pm 0.1$.

Because the error in zero level is correlated over all wavelength bins,
it represents the dominant source of uncertainty in the determination
of the continuum limits.  The contribution of this error is linear in the
number of bins included, whereas the errors due to count statistics tend
toward the square root of the number of bins.  The best determinations of
continuum level were $-1.07\pm0.69 \times 10^3 \cu$ between 977 and 1020
\angstroms and $-3.2\pm 3.8 \times 10^2 \cu$ between 1028 and 1057 \angstroms.
These measurements place 2$\sigma$ upper limits to the continuum in these 
spectral ranges of 310 and 440 \cu respectively.

\section{Comparison of results with models of the FUV background}

\subsection{Isothermal models}

The simplest (and most often used) model of emission from the warm
and hot phases of the interstellar medium is that of an optically thin, isothermal,
collisionally excited plasma.  This simple model does not take into
account how the plasma was heated, nor any temperature variations within
the plasma as it cools.  For an emission line from a specific transition $(k\rightarrow j)$
from an ion $i$ the emitted intensity $I_{kj}$ is:
\begin{equation}
I_{kj}= \int_0^\infty \int_0^\infty j_{\nu} d\nu dx = 
           h\nu_{kj} \int_0^\infty n_i n_e \gamma_{kj} dx \label{eq1}
\end{equation}
where $n_i$ is the density of the ion $i$ in the plasma and $n_e$ is the electron
density. The quantity $\gamma_{jk}$ represents the collision strength into
level $k$ and the probability of decay into level $j$.  Determining this
quantity is a complicated quantum mechanical calculation and is the subject
of numerous papers.  For the purposes of this discussion values from Landini
and Monsignori-Fossi (1990) have been used. For an isothermal plasma
equation~\ref{eq1} can be further reduced to:
\begin{equation}
I_{kj}= h\nu_{kj} {{n_i}\over{n_{\rm HI}}} {{n_{\rm HI}}\over{n_e}} \gamma_{jk} \int_0^\infty n_e^2 dx 
\end{equation}
The quantities $\frac{n_i}{n_{\rm HI}}$ and $\frac{n_{\rm HI}}{n_e}$ can be determined
from thermodynamics by assuming a temperature, density, and elemental abundances for
the plasma.  
We use the abundances of
Anders and Grevesse (1989).
If the intensity $I_{kj}$ is measured and a temperature is assumed,
only the value of the integral (known as the emission measure) is left
as a free parameter.
In practice, ratios of two emission lines, preferably from the same ion,
are used to fix the temperature.  If the temperature cannot be determined
from observations, the emission measure is usually given as the locus of
points over a range of temperatures that could produce the observed emission.

Using the upper limits obtained by the DUVE instrument, we can place upper
limits to the emission measure of isothermal plasma models for the interstellar
medium.  
In the case of local emission, such as would be expected to be associated
with the gas in the local bubble, we would expect  minimal dust absorption
between us and the emitting gas.  In this case, we can use the DUVE 
emission limits 
directly to obtain upper limits to the local emission measure.
\ifpp{
\begin{figure}
\begin{center}
\ \psfig{file=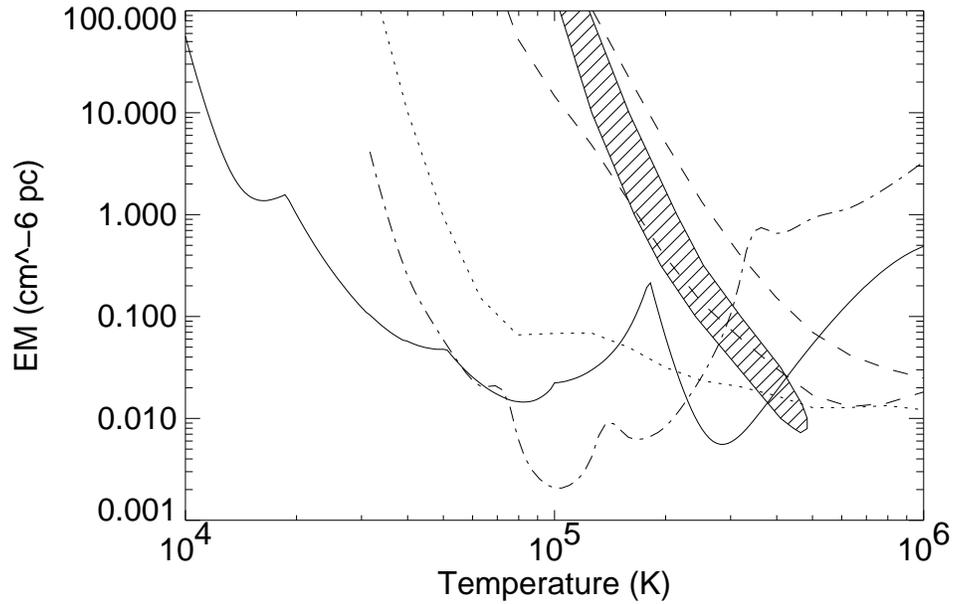}
\caption{
Upper limit local emission measures
derived from this work are shown as a solid line.  The dashed lines represent
X-ray emission measure limits derived for Wisconsin B and C band rocket borne
observations in this direction (McCammon \etal 1983).  The dot-dashed line represents emission measure
limits determined from UVX (1400 to 1800 \angstrom) observations by Martin and
Bowyer (1990).  The dotted line represents upper limits determined from EUVE
observations
by Jelinsky {\em et al} (1995).  The hatched area is the parameter space cited by
Paresce and Stern (1981) as being the allowed region to create observed broadband
EUV and Soft X-ray emission with a single temperature plasma. 
\label{fig14}
}
\end{center}
\end{figure}
}

The emission measure of the interstellar medium is constrained by upper limits
to five emission lines in this band.  Although our observations were limited to
 $\sim 7500$ square degrees of sky, we assume that the results are global
in nature.  Although this is certainly not the case in detail, it is the
only option we have given the limited number of ISM emission measurements.
With this assumption we compare our results to the Martin and Bowyer results 
for \CIV which sampled much
smaller sky areas in several different directions.  It should be noted that
Martin and Bowyer detected \CIV emission along every high galactic latitude line of sight they observed.  

The lines that limit the emission measure are:
\begin{enumerate}
\item \MgII \singlet 1027, which peaks at about $1.5\tten{4}$ K, constrains the
emission measure at temperatures below $2.0\tten{4}$ K.  Because of the nearby
interference from \HI \singlet 1025 emission, the observational limit obtained is high.
\item \CII \singlet 1037, which peaks at a temperature of about 5\tten{4} K,
provides the best constraint for temperatures between $2.0\tten{4}$ and
$5.1\tten{4}$ K.  Because our simple model does not include photoionization,
this may understate the constraint, as \CII is expected to be produced 
by photoionization due to FUV radiation from stars.
\item \CIII \singlet 977, which  peaks at $8.4\tten{4}$ K, provides the best constraint
between $5.1\tten{4}$ and $1.0\tten{5}$ K.
\item \NIII \singlet 991, which peaks at 1\tten{5} K, constrains the emission
measure between 1.0\tten{5} and 1.8\tten{5} K.
\item \OVI \doublet 1032,1038, which peaks at 2.8\tten{5} K, provides
upper limits to the emission measure above 1.8\tten{5} K.
\end{enumerate}
Figure~\ref{fig14}\ shows these emission measure upper limits.  In this figure,
the solid line represents limits to local emission measure derived from this work.
The dashed lines represent emission measure limits determined by rocket borne
broad band X-ray observations in the same direction as the DUVE observations
(\cite{mccammon}).
The dot-dashed line represents limits to emission measure determined by
\CIV \doublet 1548,1556 observations by Martin and Bowyer (1990).  The
dotted line represents upper limits determined by Jelinsky \etal (1995)
from
EUVE observations.  The hatched area is the parameter space cited by 
Paresce and Stern (1981) as being the region allowed if broadband
EUV and soft X-ray emission are produced by a single temperature plasma.

\ifpp{
\begin{figure}
\begin{center}
\ \psfig{file=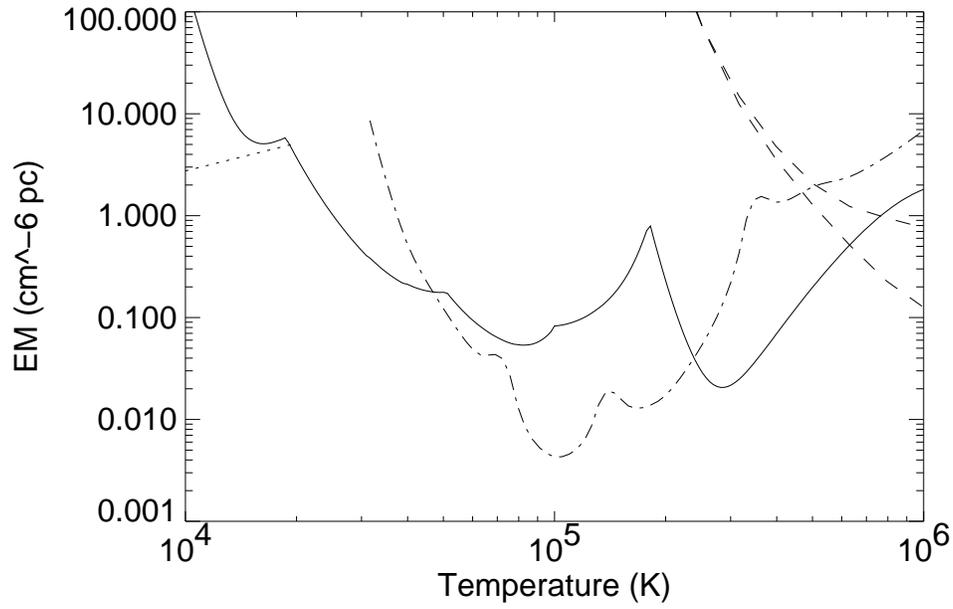}
\caption{
Upper limits to the emission
measure of the galactic halo as derived by this work are shown as a solid
line.  The dot-dashed line represents limits to the halo emission as determined by
Martin and Bowyer (1990) from UVX observations.  The dashed lines represent B and C  
band observations by McCammon {\em et al} (1983).  
The dotted line represents emission
measure limits of the warm ionized ISM as determined from optical observations 
by Reynolds (1991). 
\label{fig15}}
\end{center}
\end{figure}
}
In the case where the emission is due to non-local gas, such as emission
due to a hot galactic halo, we must consider absorption by intervening
dust.  Using the FUV extinction curve published by Sasseen {\em et al} 
(1996)
and a standard gas to dust ratio, we are able to estimate the optical depth 
of dust per unit hydrogen column, or equivalent cross section.  This equivalent
cross section varies
nearly linearly between 3.4\tten{-21} cm$^{2}$ at 950 \angstrom to 
2.6\tten{-21} cm$^{2}$ at 1080 \angstrom.  Along the lines of sight of the
DUVE observations, the mean galactic \HI column density is 4.5\tten{20} cm$^{-2}$
resulting in optical depths of between 1.2 and 1.5 and extinctions (1-$e^{-\tau}$) of 72\% at 1032 \angstroms.  This raises the 
emission
measure limits by a factor of ${e^{\tau}}$, which is between 3.3 and 4.1 over
our wavelength range.  
These upper limits to halo
emission measure are shown in Figure~\ref{fig15}.  As with the previous figure,
the dashed lines represent the Wisconsin soft X-ray observations and the
dot dashed lines represent the Martin and Bowyer FUV observations.  Both 
these measurements were corrected for dust absorption associated with
a hydrogen column of 4.5\tten{20} \mbox{\percmsqr.} 
The dotted line is the emission measure determined from optical
observations by Reynolds (1991) for the warm ionized phase of the ISM.

The points where the lines
representing various determinations of these limits cross
can be used to provide limits to the
temperature of the isothermal models.  The results  in
Figures~\ref{fig14} and~\ref{fig15} 
show that the DUVE observations limit the temperature of an isothermal
model of the emission observed by Martin and Bowyer to between 5\tten{4}
and 2.2\tten{5} K.  The temperature of an isothermal
model of the gas observed in the soft X-rays must be at a temperature above 
5\tten{5} K.

\subsection{Evolutionary Models}
\label{code}
 
Following Edgar and Chevalier (1986), we examine line emission from the hot 
interstellar medium in a simple evolutionary
model by considering
a constant  source of gas at
$T_{\rm hot}$ and a sink of gas at $T_{\rm cold}$; we assume that the mass 
flow rates at each temperature 
are equal and constant.  
The emergent
intensity of a specific line in such a model is:
\begin{equation}
I_{kj} =  h\nu_{kj} 
  \int_{T_{\rm cold}}^{T_{\rm hot}} 
  \frac{n_{\rm OVI}}{n_{\rm HI}} 
  \frac{n_{\rm HI}}{n_e} 
  \gamma_{kj}(T) {{d(EM)}\over{dT}} dT 
\end{equation}
where EM has its usual definition.
This can be simplified by making assumptions about the pressure/density evolution
(isobaric or isochoric) and the ionization state of the material involved.
Even without such assumptions, the intensity can be calculated numerically by
assuming a pressure, density, and temperature distribution.

Because of the importance of nonequilibrium ionization effects in low
density collisionally ionized gas at the temperature range in question, the assumption of ionization
equilibrium may not hold in many cases.  Solving for the nonequilibrium
ionization state in the absence of photoionization requires solving a 
system of linear differential equations of the form
\begin{equation}
\deriv{X_i}{t}=n_e \left[ C^{ion}_{i-1} X_{i-1} - 
(C^{ion}_i+C^{rec}_i) X_i + C^{rec}_{i+1} X_{i+1} \right] 
\end{equation}
where $X_i$ represents the fractional abundance of element $X$ in ionization
state $i$.  The $C^{ion}_i$ terms represent all collisional processes leading 
to ionization from state $i$ to state $i+1$.  The $C^{rec}_i$ terms represent 
all processes leading to recombination from state $i+1$ to state $i$.   These
rate coefficients can further be subdivided into
\begin{equation}
C^{ion}_i = C_{elec}^{ion}(i,T) + C^{ion}_{auto}(i,T) + C^{ion}_{c-t}(i,T) 
\end{equation}
\begin{equation}
C^{rec}_i = C_{rad}^{rec}(i,T) + C^{rec}_{2-e}(i,T) + C^{rec}_{c-t}(i,T).  \label{eq2}
\end{equation}
The rate coefficient for ionization by collisions with electrons is $C^{ion}_{elec}$,
$C^{ion}_{auto}$ is the rate coefficient for excitation autoionization, 
$C_{rad}^{rec}$ is the rate coefficient for radiative recombination, and
$C_{2-e}^{rec}$ is the rate coefficient for dielectronic recombination.
The $C^{ion}_{c-t}$ and $C^{rec}_{c-t}$ terms represent rates for charge transfer
ionization and recombination.  Because these rates are in general important
only for charge transfers from hydrogen and helium atoms to more highly 
ionized atoms, 
and because hydrogen
and helium are fully ionized in the temperature range of interest, we will
ignore these terms.
The approximations for these coefficients were taken from Arnaud and
Rothenflug (1985), and Shull and Van Steenberg (1982).

By assuming a fully ionized plasma of 90\% hydrogen and 10\% helium as the
source of the electron density ($n=1.91 n_e$), we can solve the set of 
equations for each
element individually.  Because the set of equations is stiffly coupled and
therefore subject to numerical instability, we integrated them using an 
implicit multistep stiffly stable method described by Gear (1971).

We determined a cooling rate 
from a radiative cooling curve $\Lambda(T)$ and assumed 
density structure $n(t)$. 
\begin{equation}
\deriv{T}{t}=-\frac{2\Lambda(T)}{3nk}+\frac{2T}{3n}\deriv{n}{T}  
\end{equation}
\begin{equation}
\deriv{T}{t}=\left\{ \begin{array}{ll}
      -\frac{2\Lambda(T)}{3nk} & \mbox{\rm isochoric\ cooling} \\
      -\frac{2\Lambda(T)}{5nk} & \mbox{\rm isobaric\ cooling}
      \end{array} \right. 
\label{eq3}
\end{equation}
Equations~\ref{eq2}--\ref{eq3} thus become
\begin{equation}
\deriv{X_i}{T}={{\deriv{X_i}{t}}\over{\deriv{T}{t}}}= \frac{K n_e^2 k}{\Lambda(T)}
   \left[ C^{ion}_{i-1} X_{i-1} - 
(C^{ion}_i+C^{rec}_i) X_i + C^{rec}_{i+1} X_{i+1} \right] 
\end{equation}
where $K$=2.87 for isochoric cooling and $K$=4.78 for isobaric cooling.

The initial conditions for the problem were considered to be the equilibrium
ionization values at the starting temperature ($T_{\rm hot}=10^{7}$ K).  The quantity $\Lambda(T)$ was calculated 
using the emission code of Landini and Monsignori Fossi (1990) using
our calculated ionization fraction.
New ionization fractions for important elements were calculated numerically 
using the integration code  described above.  The emission
was then recalculated using the new ionization fractions and
summed over all elements to obtain a new $\Lambda(T)$.
The procedure was iterated until $\Lambda(T)$ converged.

In this model we assumed the dominant cooling mechanism was radiative
cooling.  Thus the time a parcel of material spends in a given temperature
range, and therefore the total amount of material in that range, is inversely 
proportional to the rate of cooling of that material.  Following standard
practice we considered
three density/pressure constraints: 1) isobaric evolution wherein all the
gas is in pressure equilibrium, 2) isochoric evolution in which all the gas is
at a constant density  and 3) a mixed model in which the gas initially evolves 
isochorically
and then at some temperature begins to evolve isobarically. 
Using the value of the intensity of \CIV emission as determined by 
Martin and Bowyer,
we can calculate the expected intensity at \OVI \doublet 1032,1038.
In the case of an isobaric model, an \OVI intensity of 1.9\tten{4} \pu is
predicted, which is a factor of 2.5 greater than our observed upper limit.  In the case of 
an isochoric model, the predicted intensity is 5\tten{3}, 
which is somewhat
below the upper limit.  Mixed models fall between these values depending on the
temperature of the isochoric-isobaric transition.  These models will be
discussed further in Section~\ref{fountain}. 

\subsection{The Smith and Cox model}

Smith and Cox (1974) developed a model of the ISM
based on an estimate of the volume of the interstellar
medium occupied by supernova remnants.  They 
considered the expansion of isolated supernova
remnants with energy 4\tten{50} erg in an ambient medium of density
1.0 cm$^{-3}$.  This calculation resulted in remnants of radius $\sim 40$ pc
which persist for 4\tten{6} years.  They calculated a ``porosity''
of the ISM, $q=r \tau V_{\rm SNR}$ where $r$ is the average supernova
rate per unit volume, $\tau$ is the lifetime of an isolated SNR,
and $V_{\rm SNR}$ is the time averaged volume of the SNR over this
lifetime.  Their calculation indicated that the value of $q$ was near
0.1.  They also estimated that the probability of intersection
of a SNR with other remnants was about 50\%, and showed these interactions
were an important key to the state of the ISM.

Slavin and Cox (1992, 1993) revised the Smith and Cox model to describe
the spherically symmetric expansion of a supernova remnant 
into a warm medium of density 0.2 \percmcubed and temperature $10^4$ K.
Their model includes a magnetic
field in the ambient medium which serves to apply an additional nonthermal
pressure against the expansion.
They assumed that compression and expansion of the matter in the remnant
results in a magnetic field that is proportional to the mass density, which
is valid for plane parallel shocks.  
Their model assumed a magnetic
field in the ambient medium of 5 \microgauss.  Because this is a one dimensional
model, we would expect these calculations to be valid only along the
magnetic equator of a supernova remnant.

The result of their simulation was a bubble of hot gas which reaches its maximum
radius of $\sim 55 \parsec$ after about 1\tten{6} years.  Following this,
the bubble slowly collapses on a time scale of $\sim 5.5\tten{6}$ years.
Surprisingly, during the collapse, the central temperature stays roughly
constant at about 3\tten{5} K, while the density inside the remnant increases.

In extending their model from an individual supernova to a model of
the interstellar medium, they assumed supernovae distributed randomly throughout
the disk with an exponential scale height of 300 \parsec.  They determined
the dependence of the maximum radius, expansion time, and collapse time
with variations in the magnetic field, gas density, and supernova energy.
They used these to determine a global average time integrated volume 
for a supernova, and used this value to determine the volume filling 
factor of supernova remnants. They found a porosity $q \leq 0.18$.  
Even with this higher limit for the
filling factor, their model shows fewer interactions between remnants than
would be expected from the earlier model of Smith and Cox.
The mean free path between bubbles and \OVI\ column density per bubble
in this model closely matches that determined from the analysis of 
Copernicus observations by Shelton and Cox (1994).

For the purpose of comparing FUV emission measurements to  a Slavin and Cox 
model 
it was necessary to simplify the model to some degree.
Rather than duplicate their entire MHD calculation, we digitized
several figures from their articles containing the density, temperature,
velocity, and total pressure profiles of the remnant at various times.
We were able to fit these curves with Chebyshev polynomials with time varying
coefficients to produce an approximation of the profiles.  
Using these density, temperature, pressure, and velocity profiles we were
able to calculate the non-equilibrium ionization and the corresponding
line emission for a ``standard" supernova remnant using the method described
in Section~\ref{code}.  Our simplified model closely matches the 
results obtained by Slavin and Cox.

We next modeled the appearance of the sky. This required several assumptions.
{Slavin and Cox assumed a constant
midplane supernova rate of 0.4\tten{-13} \parsec$^{-3}$ yr$^{-1}$,
and a supernova scale height of 300 pc.}
{For the purpose of modeling sky coverage, they assumed that all supernovae are ``standard'' supernovae
of energy 5\tten{50} ergs,
occurring in a medium of density 0.2 \percmcubed, temperature of 10000 K, 
and magnetic field of
5 \microgauss.}
{We modeled limb brightening effects in the remnant by assuming the
ions were uniformly distributed within a spherical shell with inner and outer
radii determined by the radii where $n_i n_e = {\rm max}(n_i n_e)/2$.}
{Absorption by interstellar dust was modeled by assuming a constant gas/dust
ratio. We considered an \HI distribution with three components: a Gaussian
component of scale length 110 parsecs with central density 0.39 \percmcubed,
a Gaussian component of scale length 260 parsecs and central density
0.11 \percmcubed, and an exponential component of scale height 400 parsecs
and a central density of 0.06 \percmcubed. 
The distribution of H$_2$ was modeled as an exponential distribution
with a central density of 1.333 \percmcubed and with scale height 55 \parsec
(\cite{dickey,dame,savage95}).}
{Interstellar opacities at 1032 \angstrom and 1550 \angstrom were
determined with reference to Sasseen \etal (1996) and correspond
to cross sections of 2.9\tten{-21} \cmsquared N$_{\rm H}^{-1}$ and
1.5\tten{-21} \cmsquared N$_{\rm H}^{-1}$ respectively.}
{The distance to the remnants and the associated  hydrogen column were assumed to be constant
across the face of each remnant.}
{A small component due to the local bubble was modeled as emission
from a SNR with central temperature 2\tten{6} K, and with a temperature 
gradient
as specified by R.~Smith (1996).}

Emission for both \OVI \doublet 1032,1038 and \CIV \doublet 1548,1551 were
calculated.  At both wavelengths the sky is faint,
with the median sky intensities being $<\iovi>$ = 120 \pu and
$<\iciv>$ = 350 \pu.  Based on these calculations we estimate that 
less than 10\% of the sky emits at
the level of \CIV seen by Martin and Bowyer, 5000 \pu.  
In addition to the difference
in overall intensity, there is a difference in the distribution of the
emission.  The \CIV intensity measured by Martin and Bowyer is at least
a factor of two brighter near the galactic poles than near the galactic
plane, whereas in the Slavin and Cox model, the majority of the emission
is seen near the galactic plane.

This result seems to be fairly invariant to the major parameters of the model.
Introducing a variation evolution of the remnants with altitude above the
galactic plane, $z$, could increase the fractional sky coverage of
remnants at high latitude. However, the intensity of a remnant is lower 
due to the lower densities, resulting in a longer lasting remnant.  We
expect these variations to come near to canceling, with the average sky
intensity staying roughly constant.  This effect should be included in future
models and is currently the subject of study. (\cite{sheltonth})  

Spatially and temporally correlated supernovae could also have an effect.
Slavin and Cox's calculations show a rough $\int A dt \propto E$
dependence at constant pressure parameters, thus we would expect the 
time integrated sky coverage of a single
multiple event remnant to be proportional to the number of supernovae that
occurred in the remnant, $N$.  On the other hand, the number of such remnants
present in the sky would be proportional to $\frac{1}{N}$.  Therefore we
expect the sky coverage of remnants to be roughly constant.  The remnant 
lifetime is dependent on the radiative properties of the remnant.  A rough
estimate of the remnant luminosity can be made by assuming the energy radiated
is proportional to the total energy of the supernovae, $N\times E_{sn}$, 
radiated through $\int A dt$ surface area and time.  Again the factors
cancel leaving the time averaged intensity unchanged.

\subsection{The McKee-Ostriker model}

McKee and Ostriker (1977, hereafter MO) presented a detailed model
of the interstellar medium.
The model assumed heating of the hot
medium by supernovae, and cooling of the hot medium by radiation 
and with thermal pressure equilibrium between the warm clouds and the
hot medium.
By assuming a lower pressure and
density in the ambient medium than that assumed by Smith and Cox, and
a higher energy per supernova, they
calculated a porosity $q$ which could exceed unity, rendering simple
calculations of porosity invalid due to interaction of remnants. 
Therefore, they simply assumed that the filling factor of the hot gas was large.
Using this, they devised a 3 phase model of
the ISM, consisting, on average of warm ($10^{4}$ K) clouds (some with cool ($10^2$ K)
cores) existing in pressure equilibrium with a hot ($10^{5.7-6.0}$ K)
medium at pressure $10^{3.2}$ \percmcubed K, and with an intermediate temperature evaporative 
region between the hot and warm components.  

In principle, a simple constraint which could be placed upon the MO model is 
that
of the filling factor of the hot medium, $f_{\rm hot}$.  
The \OVI emission measure
that would be expected to be present along the DUVE line of sight is:
\begin{equation}
{\rm EM} = 5.7\tten{-4} 
\left[ \frac{P}{2500 \pressure} \right]^2
   \left[ \frac{T_{\rm hot}}{5\tten{5} K} \right]^{-5.2}
   \left[ \frac{f_{\rm hot}}{0.6} \right]
   \left[ \frac{h}{1 \kpc} \right] \emunits 
   \end{equation}
Unfortunately, this emission is less than the DUVE lower limit even with filling
factors of 1. This shows that detecting
the hot medium itself is a difficult task at FUV wavelengths.  This is
because the bulk of 
FUV emission
in a MO model is produced by the interfaces between the warm and hot media.

To more accurately calculate expected emission from interfaces in a MO model, 
we used an analytic solution for conductive cloud
boundaries as described by Dalton and Balbus (1993). 
This model describes both saturated and unsaturated evaporation
of spherical clouds.  Saturated evaporation occurs when the
temperature gradients become large compared to the mean free
path of electrons in the gas.  This results in a breakdown in the
local heat transport; energy is transported many temperature
scale heights. The result is that the diffusive approximation ($q=-\kappa \del
T$) is no longer valid, resulting in much lower heat flux than would be
obtained by using the gradient of the temperature.
The form of the solutions for density, temperature, 
and velocity profiles when parameterized to a
unitless radius $y=\frac{r}{R_{\rm cloud}}$ and temperature 
$\tau=\frac{T}{T_{\rm hot}}$ has only one free parameter, the
saturation parameter $\sigma_o$, which can be approximated as the ratio of the
electron mean free path in the hot gas and the cloud radius.

Because the MO model is at pressure equilibrium, the cloud radius is
determined by the mean \HI column density of the cloud.  The model also provides velocities
of the evaporating gas, which allow calculation of non-ionization
equilibrium effects.
It is assumed that the temperature distribution in this model
is determined only by the gas flow and heat conduction, with no radiative 
losses. 
Because the gas flows toward regions of rapidly
decreasing density, the high stage ions can persist much longer and to much 
greater
radii than would be expected from collisional ionization equilibrium 
calculations.  

We calculated the emission from evaporative cloud boundaries for clouds of varying radii 
at the ``preferred'' MO values of the temperature (5\tten{5} K)
and filling factor (0.6) and a variety of pressures.
The resulting emergent intensities from the conductive boundaries were
well fit by a power law. 
Given a power law distribution of clouds with
slope ($\alpha$) and summed column density, N$_{\rm HI}$, we can calculate
the intensity of emission along the line of sight of the DUVE experiment.
Using the power law described by Dickey and Lockman (1990)
($\alpha=-4.3$)
for clouds with column density between 1.5\tten{18} and 1.2\tten{19}, we
calculate a total emission integrated along the line of sight of 
\begin{equation}
I_{\rm OVI}=1600 \left( \frac{P}{2500 \pressure} \right)^2 \left(1-e^{\frac{-N_{18}}{350}}\right) 
\end{equation}
\begin{equation}
I_{\rm CIV}=5300 \left( \frac{P}{2500 \pressure} \right)^2 
\left(1-e^{\frac{-N_{18}}{650}}\right) 
\end{equation}
where $N_{18}$ is the total line of sight hydrogen column in units 
of 10$^{18}$ \percmsqr.
Along the DUVE line of sight, this would result in \OVI and \CIV emission 
of 1200 and 2700
\pu respectively.  

Because this model depends upon the distribution of cloud sizes, which in turn
effects the number of cloud interfaces along the line of sight, it is fairly 
sensitive to the power of the cloud size distribution.  An increase in the
power law slope from -4.3 to -3.3, causes a reduction in
the number of small clouds.  This results in an intensity reduction of a 
factor of two.  Non spherical clouds, especially those with correlated
orientations, or the inclusion of magnetic fields could also change these
results.

\ifpp{
\begin{figure}
\begin{center}
\ \psfig{figure=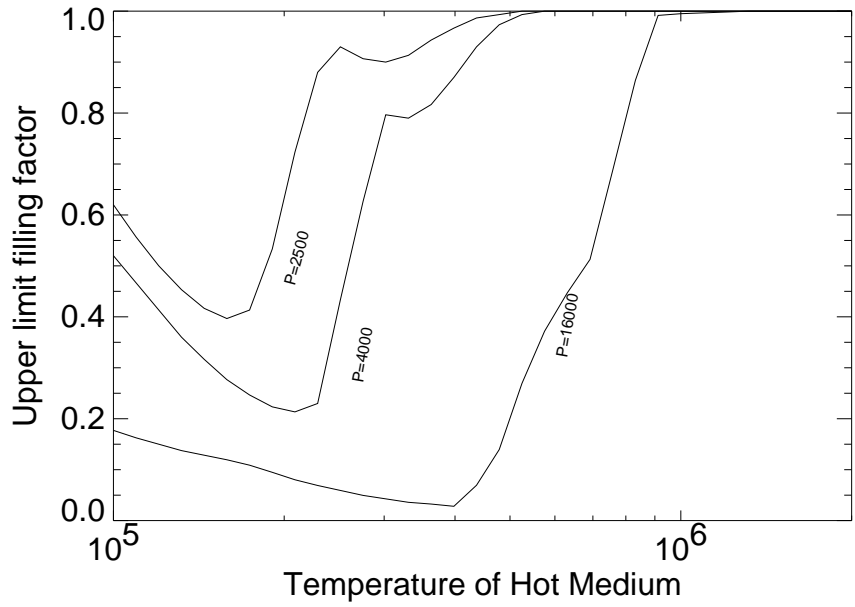}
\caption{
The solid lines show the limits to filling factor in an McKee-Ostriker model placed by the 
DUVE data as a function of temperature for 3 assumed pressures.
Pressures
shown are 2500, 4000, and 16000 \pressure.
\label{fig20}}
\end{center}
\end{figure}
}
We repeated this calculation for a wide range of temperatures
and filling factors.  
By assuming a given pressure, we can determine the locus of values of filling factor
and temperature that would produce emission at the DUVE \OVI upper limit or
\CIV emission at the level of  the Martin and Bowyer measurement.  Such curves are
plotted for 3 different pressures ( 2500, 4000, 16000 
\pressure) in 
Figure~\ref{fig20}.  The area below these curves represent allowed values
of the temperature and filling factor.

\subsection{Galactic fountain models}
\label{fountain}

Shapiro and Field (1976) first described a 
``galactic fountain'' model
of the hot interstellar medium.  They argued that the long 
cooling time of hot, low-density gas ($t \gt 10^6$ yrs) would allow time
for the hot gas to rise buoyantly above the galactic plane.  This gas would
eventually cool radiatively, condense into clouds, and fall back to the
plane.  They argued that this model could explain both the presence of hot
gas and infalling high velocity clouds.

Basic galactic fountain models follow the thermal evolution of hot
gas.   Most of these basic models assume that the gas starts its cooling
at a constant density and then, at some point, it transitions to evolution at 
constant pressure.  If we assume that the
\CIV emission observed by Martin and Bowyer is emission from a galactic
fountain, we can use these simple models to determine what the expected
\OVI intensity would be.  These estimates were produced using the non-equilibrium
ionization code described in Section~\ref{code}.  We again calculated the time
integrated emission for a cooling gas parcel for both isobaric and isochoric
cooling.  We repeated this calculation over a wide range of initial
temperatures from 10$^5$ to 10$^7$ K.  We also repeated the calculations
for models including an isochoric$\rightarrow$isobaric transition for
transition temperatures ranging from 10$^5$ to 10$^7$ K.  The values
of \CIV and \OVI intensity
were corrected for differential absorption by dust associated with a
hydrogen column of 4.5\tten{20} \mbox{\percmsqr.}

\ifpp{
\begin{figure}
\begin{center}
\ \psfig{file=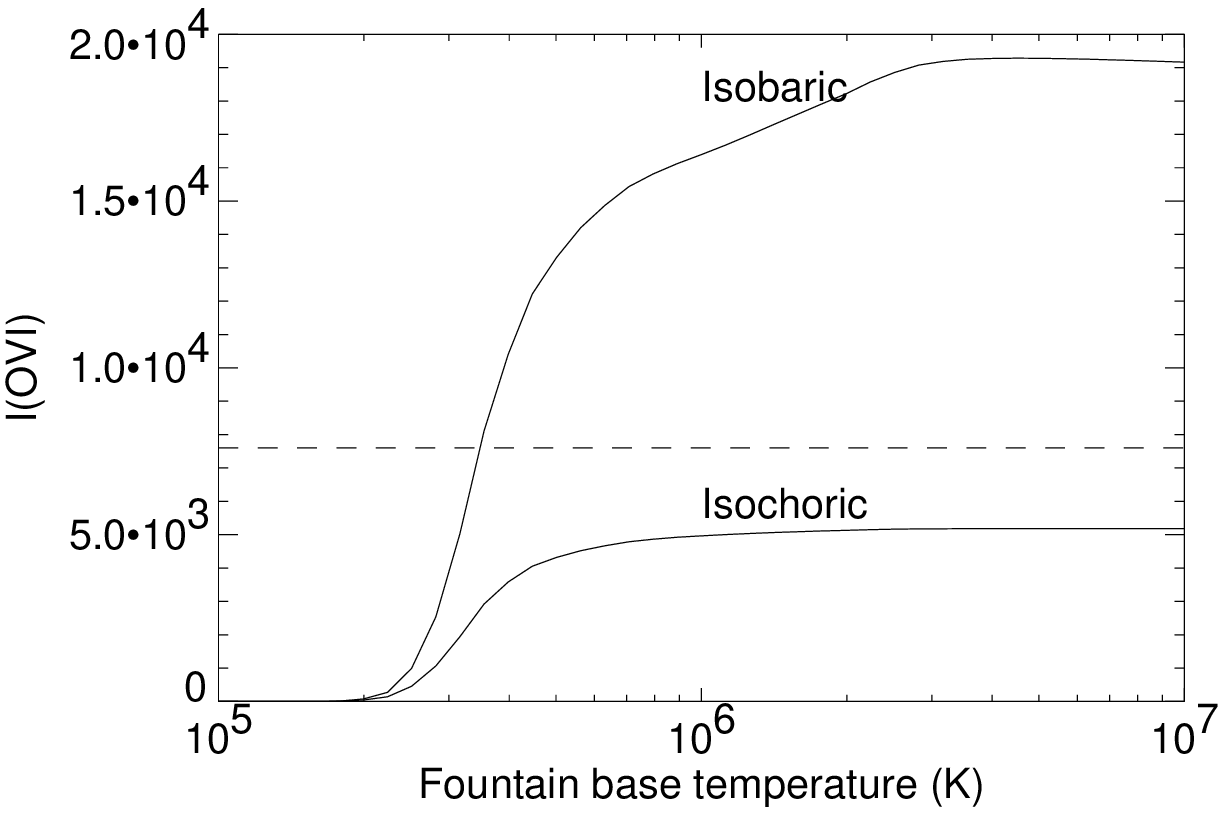}
\caption{
\OVI emission vs base temperature for simple galactic fountain models.
Both isobaric and isochoric models are shown.  The plot assumes 5000
\pu of measured \CIV emission and dust absorption associated with an \HI 
column of 4.5\tten{20}
\percmsqr.  The dotted line represents the DUVE upper limit to \OVI emission.
\label{fig21}}
\end{center}
\end{figure}
\begin{figure}
\begin{center}
\ \psfig{file=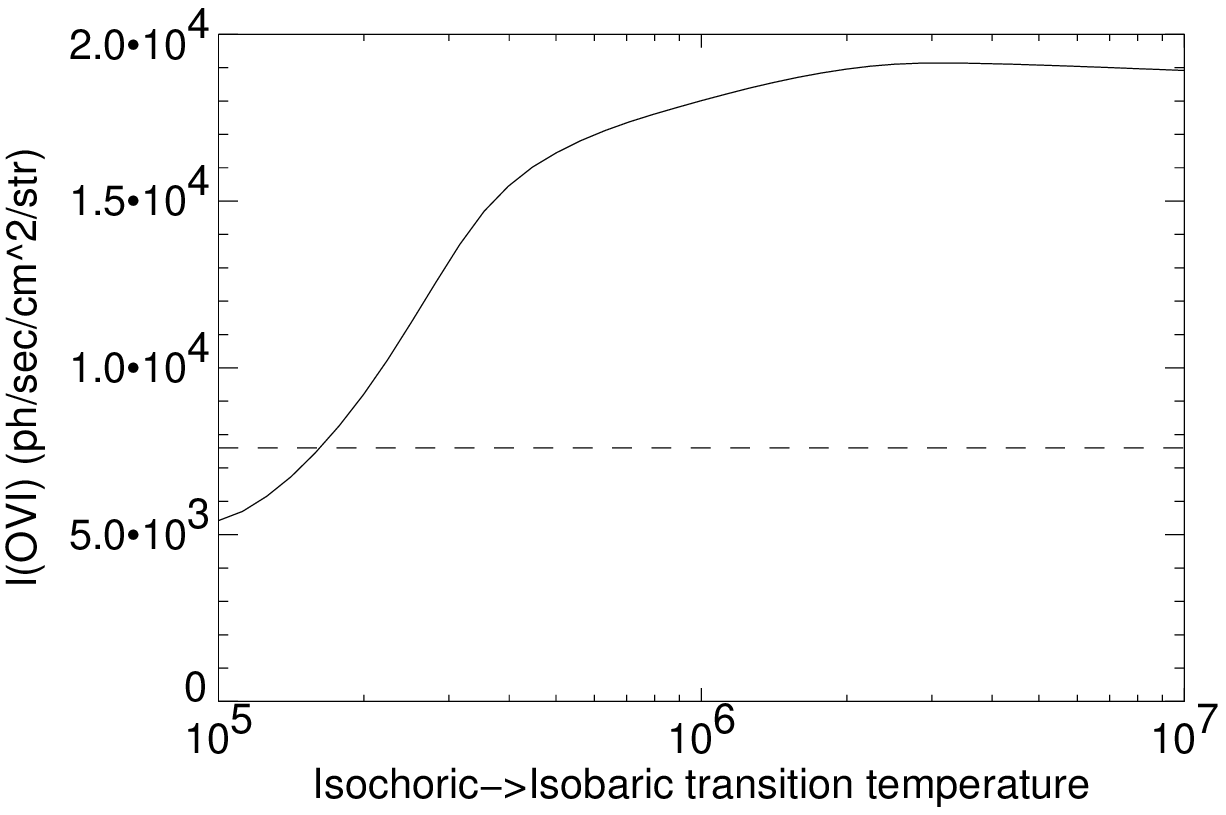}
\caption{
\OVI emission vs transition temperature for simple galactic fountain models. The
lower plot assumes 5000 \pu of \CIV emission and dust absorption associated with an
\HI column of
4.5\tten{20} \percmsqr. The dotted line represents the DUVE upper limit to
\OVI emission.
\label{fig22}}
\end{center}
\end{figure}
}
Figure~\ref{fig21}\ shows calculations of \OVI emission vs the base temperature
for simple isochoric and isobaric galactic fountain models scaled to
the Martin and Bowyer \CIV emission value. 
We can place an upper limit input temperature of 3.5\tten{5} K on
fully isobaric fountains.  Fully isochoric fountains would produce \OVI
emission at levels below the DUVE detection threshold.
In Figure~\ref{fig22}, we show models containing an isochoric/isobaric transition. 
We can limit the transition temperatures to less than 2\tten{5} K.

Many far more complex models of galactic fountains have been published
including effects outside of the realm of the simple model described above.
They include such things as non-equilibrium ionization, turbulent
mixing of hot and cold gas, and photoionization by the emitted EUV and 
X-ray radiation.  
However, a prediction
of the ratio of $\frac{\rm I(OVI)}{\rm I(CIV)}$ and the intensity of \OVI emission
per solar mass of circulation within the fountain can be obtained for
each of these models.  These predictions assume that the lines of sight
observed by DUVE and UVX can be extrapolated to give a global average 
intensity across
the entire galactic disk.  The large
solid angle observed by DUVE, (18\% of the sky), makes it likely that this
method of averaging is applicable.  While this method of averaging is
less applicable to the Martin and Bowyer \CIV measurements with their
smaller sampling solid angle, the fact that \CIV was detected along every 
high-latitude line of sight could indicate that their measurements are
typical of high latitude \CIV emission.

The model by Edgar and Chevalier (1986) includes non-equilibrium effects
at temperatures below 10$^6$ K, and transitions between isochoric and isobaric
evolution.  They predict an emission at \CIV \doublet 1549 of 560-890 \pu for
a mass flow of 4 M$_\odot$ yr$^{-1}$ and \OVI \doublet 1032,1038 emission of
1600-2400 \pu.  We can limit the mass flow rate in their model
to less than 4.75 M$_\odot$ yr$^{-1}$.  Their model also predicts 
$\frac{\rm I(OVI)}{\rm I(CIV)} = 7.2-10.8$, a much higher value than our observed limit
of $\sim$1.5.  Again these are averages over the galactic disk and could be
much brighter in some directions.

Houck and Bregman (1990) attempt to fit the observed 
scale height
and velocity distribution of
neutral gas clouds in the galaxy with a galactic fountain model. 
Their model does not include non-equilibrium ionization effects, which
they concluded would have little effect on their results.
Their results were calculated along a one dimensional grid. Their best
fit to the distribution of neutral cloud velocities was a model with 
a base temperature of 3\tten{5} K,
a base pressure of 300 \pressure, and a mass flow rate of 0.2 M$_\odot$ yr$^{-1}$.  
Although Houck and Bregman did not include explicit predictions of FUV
emission, we were able to use data from their paper to produce conservative
lower limits to the equilibrium \CIV and \OVI intensities of this model.  
By digitizing figures from their paper, we were able to estimate $I_{\rm OVI}$ to be approximately 1200 \pu per solar mass of flow and 
$\frac{\rm I(OVI)}{\rm I(CIV)} =  0.3$.  Both of these values are consistent with
the DUVE observations as well as those of Martin and Bowyer.  However, because
this model is designed to match infalling clouds, changes in our understanding
of these could could greatly change the predicted intensity.

Shull and Slavin (1994) added the effects of turbulent
mixing of hot and cold gas into galactic fountain models.  In this model,
they calculate the ratio for emission $\frac{\rm I(OVI)}{\rm I(CIV)} = 1.2\pm 0.2$
for $\log T = 5.3\pm 0.3$, resulting in a predicted \OVI emission of 5000
to 7000 \pu at mass flow rates between 15 and 25 M$_\odot$ yr$^{-1}$.  These
predictions are based upon a range of mixing velocities, abundances, and mean
temperatures.  These parameters were used to determine the interstellar pressure
required to produce the observed \CIV emission. The 
$\frac{\rm I(OVI)}{\rm I(CIV)}$ ratio determined through this method is
consistent with the DUVE observations.  We are able to limit mass
flow in this model to 21 to 38 M$_\odot$ yr$^{-1}$.

\ifpp{
\begin{table}[t]
\begin{center}
\begin{tabular}{lccc}
\hline
Model & Edgar \& & Houck \& & Shull \& \\
           & Chevalier 1986 & Bregman  1990 & Slavin 1994 \\
\hline     
T$_{\rm max}$ (K) & $10^6$ & 3\tten{5} & $10^{5.3\pm0.3}$ \\
$\frac{\rm I(\OVI)}{\rm \dot{M}[\odot]}$ & 1600-2400 & 1200 & 200-350 \\
$\frac{\rm I(\OVI)}{\rm I(\CIV)}$ & 7.2--10.8 & 0.3 & 1.0--1.4 \\
${\rm \dot{M}_{max} (M_\odot yr^{-1}) }$ & \lt 3.2--4.75 & \lt 6.3 & \lt 21--38 \\
\hline
\end{tabular}
\caption{Constraints to galactic fountain models.}
\label{tab5}
\end{center}
\end{table} 
}
The predicted \OVI intensities and the limits to mass flow rates in each of
these
models is shown in Table~\ref{tab5}.

\section{Discussion}

Although the DUVE data are able to place constraints to the models discussed
above, it is important to note that the model parameters are sufficiently 
flexible, the DUVE data alone cannot rule them out.  The patchy 
distribution of the emitting gas in the galactic fountain and Smith and Cox
models makes it possible, if somewhat unlikely, that the distribution of gas 
along the Martin and Bowyer lines of sight bears no resemblance to that 
sampled by the DUVE observations.  
Observations of both \CIV and \OVI emission along the same line of sight
would be extremely useful.
The predicted distribution of these emissions varies 
from highly pole brightened (galactic fountain) to fairly uniform (MO models)
to plane brightened (Smith and Cox), thus an all sky spectral survey of 
diffuse FUV emission would greatly enhance our understanding of the ISM.

\section{Conclusions}

We designed and built an instrument for the study of emission from the
diffuse ISM  at wavelengths between
950 and 1080 \angstroms.  This instrument was flown on July 24, 1992 attached
to the second stage of a Delta II launch vehicle.  It achieved orbit and operated
as planned, providing 4151 seconds of observations during orbital night
periods.

Observations made by the DUVE instrument did not detect 
emission from the hot phase of the interstellar medium.  We were, however,
able to place new upper limits to emission from \OVI \doublet 1032,1038 
, \CII \singlet 1037 , \CIII \singlet 977, and \NIII \singlet 971 which
significantly constrain the parameters of the hot ISM.  We have used these
limits to determine constraints to the emission measure of both local and
halo gas between $10^4$ and $10^6$ K.

We have modeled \OVI and \CIV emission from scattered supernovae and compared
the results to our observations.
Our model predicts emission at levels below the previously measured \CIV
values and the \OVI limits reported here.  

We have modeled \OVI and \CIV emission from clouds in a 
standard McKee Ostriker model.  The produced emission in this model
is also below the DUVE \OVI upper limits and the previously measured 
\CIV values.
We have placed limits to the filling factor of the hot medium versus  its
temperature.

We are able to place limits to the base temperatures and isochoric
transition temperatures in simple galactic fountain models.
Our observations are inconsistent with the I[\OVI]/I[\CIV] ratios
predicted by simple galactic fountain models with base gas temperatures of above about 
$10^{5.6} $K. Our observations are consistent
with galactic fountain models with lower temperatures. 
Our observations place limits to the mass flow rates of several 
galactic fountain models.

\section{Acknowledgements}

We would like to thank Michael Lampton for his assistance with many aspects
of this project, Charlie Gunn at NASA/OLS
for providing the opportunity for this mission, the staff of the
Space Sciences Lab and the McDonnell Douglas Delta Program Office for
their assistance.
This work has been supported by NASA grant NGR-05-003-805.

\ifms{
\clearpage
\figcaption[fig1.eps]{A schematic of the DUVE instrument. The dot-dash
lines represent the light path from entrance aperture to spectral detector.
The dotted line represents the path of zero order light from the Wadsworth
grating.\label{fig5}}
\figcaption[fig2.eps]{DUVE area solid-angle product or grasp \label{fig9}}
\figcaption[fig3.eps]{Raw spectral detector images from the DUVE instrument's 
flight. The top image is a sum of shutter open images.  The vertical stripe
is \HI \singlet 1025 airglow.  The bottom image is a sum of the corresponding
shutter closed images.  
Integration times are 1583 and 1529 seconds respectively.
\label{fig11}} 
\figcaption[fig4.eps]{The solid line represents
the flight spectral data after subtraction of the background image and 
convolution with a line spread function.
The dashed lines represent $\pm 3 \sigma$ error levels.\label{fig12}}
\figcaption[fig5.eps]{Upper limits to line emission placed 
by this work are shown
as a solid line.  Previous limits by Edelstein and Bowyer (1993) are shown by
a dashed line. Limits determined from Voyager UVS measurements 
as analyzed by Edelstein, Bowyer, and Lampton (1997) are presented 
as a dot-dash line. The diamond with error bars
is a measurement of OVI 1032,1038 \angstrom emission by Dixon {\em et al.} (1996).  The vertical lines show the positions of several important astrophysical
lines.
\label{fig13}}
\figcaption[fig6.eps]{
Upper limit local emission measures
derived from this work are shown as a solid line.  The dashed lines represent
X-ray emission measure limits derived for Wisconsin B and C band rocket borne
observations in this direction (McCammon \etal 1983).  The dot-dashed line represents emission measure
limits determined from UVX (1400 to 1800 \angstrom) observations by Martin and
Bowyer (1990).  The dotted line represents upper limits determined from EUVE
observations
by Jelinsky {\em et al} (1995).  The hatched area is the parameter space cited by
Paresce and Stern (1981) as being the allowed region to create observed broadband
EUV and Soft X-ray emission with a single temperature plasma. 
\label{fig14}
}
\figcaption[fig7.eps]{
Upper limits to the emission
measure of the galactic halo as derived by this work are shown as a solid
line.  The dot-dashed line represents limits to the halo emission as determined by
Martin and Bowyer (1990) from UVX observations.  The dashed lines represent B and C  
band observations by McCammon {\em et al} (1983).  
The dotted line represents emission
measure limits of the warm ionized ISM as determined from optical observations 
by Reynolds (1991). 
\label{fig15}}
\figcaption[fig8.eps]{
The solid lines show the limits to filling factor in an McKee-Ostriker model placed by the 
DUVE data as a function of temperature for 3 assumed pressures.
Pressures
shown are 2500, 4000, and 16000 \pressure.
\label{fig20}}
\figcaption[fig9.eps]{
\OVI emission vs base temperature for simple galactic fountain models.
Both isobaric and isochoric models are shown.  The plot assumes 5000
\pu of measured \CIV emission and dust absorption associated with an \HI 
column of 4.5\tten{20}
\percmsqr.  The dotted line represents the DUVE upper limit to \OVI emission.
\label{fig21}}
\figcaption[fig10.eps]{
\OVI emission vs transition temperature for simple galactic fountain models. The
lower plot assumes 5000 \pu of \CIV emission and dust absorption associated with an
\HI column of
4.5\tten{20} \percmsqr. The dotted line represents the DUVE upper limit to
\OVI emission.
\label{fig22}}

\clearpage

\begin{table}
\begin{minipage}{5.95truein}
\begin{center}
\begin{tabular}{crrr}
\hline
Species & \multicolumn{1}{c}{$\singlet (\angstrom)$}
        & \multicolumn{1}{c}{$I (\pu)$}
        & \multicolumn{1}{c}{$I (\intensity)$} \\  
\hline  
\hline  
\HI\footnotemark & 972 & $\leq$ 7.4\tten{4} & $\leq$ 1.5\tten{-6} \\
\setcounter{firstfn}{\value{footnote}}
\CIII & 977 & $\leq$ 4.0\tten{4} & $\leq$ 8.1\tten{-7} \\
\setcounter{secondfn}{\value{footnote}}
\OI$^{\thefirstfn}$ & 989 & $\leq$ 6.1\tten{3} & $\leq$ 1.2\tten{-7} \\
\NIII& 991 & $\leq$ 5.5\tten{3} & $\leq$ 1.1\tten{-7} \\
\SiII & 992 & $\leq$ 5.7\tten{3} & $\leq$ 1.1\tten{-7} \\
\SiIII & 996 & $\leq$ 1.6\tten{4} & $\leq$ 3.2\tten{-7} \\ 
\NeVI & 1006 & $\leq$ 1.3\tten{4} & $\leq$ 2.6\tten{-7} \\
\ArVI & 1008 & $\leq$ 1.0\tten{4} & $\leq$ 2.0\tten{-7} \\
\HI$^{\thefirstfn,}$\footnotemark & 1025 & 2.26$\pm$0.26\tten{5} & 4.38$\pm$0.49
\tten{-6} \\ 
\setcounter{thirdfn}{\value{footnote}}
\OVI\footnotemark & 1032, 1038 & $\leq$ 7.6\tten{3} & $\leq$ 1
.4\tten{-7} \\
\setcounter{fourthfn}{\value{footnote}}
\CII & 1037 & $\leq$ 3.9\tten{3} & $\leq$ 7.4\tten{-8} \\
\ArI$^{\thefirstfn}$ & 1050 & $\leq$ 6.4\tten{3} & $\leq$ 1.2\tten{-7} \\ 
\SiIV & 1067 & $\leq$ 2.9\tten{4} & $\leq$  5.4\tten{-7} \\
\SIV & 1070 & $\leq$ 5.1\tten{4} & $\leq$ 9.5\tten{-7} \\  
\hline 
\end{tabular}
\caption{Upper limits to line emission placed by the DUVE data}
\label{tab4}
\end{center}
\footnotetext{$^\thefirstfn$ Anticipated airglow line}
\footnotetext{$^\thethirdfn$ Detected at 37$\sigma$.} 
\footnotetext{$^\thefourthfn$ This limit is total emission from the doublet
based upon joint statistics
by assuming $\frac{I(1032)}{I(1038)}=2$.  Upper limits for the individual
components of the doublet are $I(1032) \leq$ 5400 \pu and $I(1038) \leq$
4400 \pu.} 
\end{minipage}
\end{table}

\begin{table}
\begin{center}
\begin{tabular}{lccc}
\hline
Model & Edgar \& & Houck \& & Shull \& \\
           & Chevalier 1986 & Bregman  1990 & Slavin 1994 \\
\hline     
T$_{\rm max}$ (K) & $10^6$ & 3\tten{5} & $10^{5.3\pm0.3}$ \\
$\frac{\rm I(OVI)}{\rm \dot{M}[\odot]}$ & 1600-2400 & 1200 & 200-350 \\
$\frac{\rm I(OVI)}{\rm I(CIV)}$ & 7.2--10.8 & 0.3 & 1.0--1.4 \\
${\rm \dot{M}_{max} (M_\odot yr^{-1}) }$ & \lt 3.2--4.75 & \lt 6.3 & \lt 21--38 \\
\hline
\end{tabular}
\caption{Constraints to galactic fountain models.}
\label{tab5}
\end{center}
\end{table} 
}

\begin{thebibliography}{}
\bibitem[Anders \& Grevesse 1989]{anders}Anders, E. \& Grevesse, N. 1989, Geochim. Cosmochim., {53}, 197.
\bibitem[Arnaud \etal 1985]{arnaud85}Arnaud, M., \& Rothenflug, R. 1985, A\&AS, {60}, 425.
\bibitem[Bowyer, Field \& Mack 1968]{bfm68}Bowyer, C.~S., Field, G.~B., \& Mack, J.~E. 1968, Nature, {351}, 32.
\bibitem[Chakrabarti \etal 1984]{chakrabarti}Chakrabarti, S., Kimble, R. \& Bowyer, S. 1984, JGR, {89}, 5660.
\bibitem[Dalton \& Balbus 1993]{dalton}Dalton, W.~W., \& Balbus, S.~A. 1993, ApJ, {404}, 625.
\bibitem[Dame \& Thaddeus 1995]{dame}Dame, T.~M., \& Thaddeus, P. 1995, ASP Conf. Series, {80}, 15.
\bibitem[Dickey \& Lockman 1990]{dickey}Dickey, J.~M., \& Lockman, F.~J. 1990, ARA\&A, {28}, 215.
\bibitem[Dixon \etal 1996]{dixon}Dixon, W.~V., Davidsen, A.~F. \& Ferguson, H.~C. 1996, ApJ, {465}, 288.
\bibitem[EBL1997]{ebl97}Edelstein, J., Bowyer, S., and Lampton, M. 1997, Submitted to ApJ.
\bibitem[Edelstein \& Bowyer 1993]{edelstein}Edelstein, J. \& Bowyer, S. 1993, AdSpR, {v13}, n12, 307.
\bibitem[Edgar \& Chevalier 1986]{edgar}Edgar, R.~J., \& Chevalier, R.~A. 1986, ApJ, {310}, L27.
\bibitem[Gear 1971]{gear}Gear, C.~W. 1971, {Numerical Initial Value Problems in Ordinary Differential Equations}, (Englewood Cliffs: Prentice Hall).
\bibitem[Holberg 1986]{holberg}Holberg, J.~B. 1986, ApJ, {311}, 969.
\bibitem[Houck \& Bregman 1990]{houck}Houck, J.~C., \& Bregman, J. 1990, ApJ, {352}, 506.
\bibitem[Hurwitz \& Bowyer 1996]{hurwitz96}Hurwitz, M., \& Bowyer S. 1996, ApJ, {465}, 296.
\bibitem[Jelinsky \etal 1995]{jelinsky}Jelinsky, P., Vallerga, J.~V. \& Edelstein, J. 1995, ApJ, {442}, 653.
\bibitem[Jenkins 1978a]{jenkins78a}Jenkins, E.~B. 1978a, ApJ, {219}, 845.
\bibitem[Jenkins 1978b]{jenkins78b}Jenkins, E.~B. 1978b, ApJ, {220}, 107.
\bibitem[Korpela 1997]{korpela97}Korpela, E.~J. 1997, Ph.~D. Thesis, University of California.
\bibitem[Landini \& Monsignori Fossi 1990]{landini}Landini, M., \& Monsignori Fossi, B.~C. 1990, A\&AS, {82}, 229.
\bibitem[Martin \& Bowyer 1990]{martin90}Martin, C. \& Bowyer, S. 1990, ApJ, {350}, 242.
\bibitem[McCammon \etal 1983]{mccammon}McCammon, D., Burrows, D.~N., Sanders, W.~T., Kraushaar, W.~L. 1983, ApJ, {269}, 107.
\bibitem[McKee \& Ostriker 1977]{mckee77}McKee, C.~F. \& Ostriker, J.~P. 1977, ApJ, {218}, 148.
\bibitem[Paresce \& Stern 1981]{paresce}Paresce, F. \& Stern, R. 1981, ApJ, {247}, 89.
\bibitem[Raymond 1992]{raymond92}Raymond, J.~C. 1992, ApJ, {384}, 502.
\bibitem[Reynolds 1991]{reynolds}Reynolds, R.~J. 1991, ApJ, {372}, L17.
\bibitem[Rosen, Bregman \& Norman 1993]{rbn93}Rosen, A., Bregman, J.~N., \& Norman, M.~L. 1993, ApJ, {413}, 137.
\bibitem[Sasseen 1996]{sasseen}Sasseen, T.~P., Hurwitz, M., Dixon, W.~V., Bowyer, S. 1996, BAAS, {188}, 07.02. 
\bibitem[Savage \& de Boer 1981]{savage81}Savage, B.~D., \& Massa, D. 1987, ApJ, {314}, 380.
\bibitem[Savage 1995]{savage95}Savage, B.~D. 1995, ASP Conf. Series, {80}, 233.
\bibitem[Sembach \etal 1995]{sembach95}Sembach, K.~R., Savage, B.~D., \& Lu, L. 1995, ApJ, {439}, 672.
\bibitem[Shapiro \& Field 1976]{shapiro76}Shapiro, P.~R. \& Field, G.~B. 1976, ApJ, {205}, 762.
\bibitem[Shapiro \& Benjamin 1991]{shapiro91}Shapiro, P.~R. \& Benjamin, R.~A. 1991, PASP, {103}, 923.
\bibitem[Shelton 1996]{sheltonth}Shelton, R.~L. 1996, Ph.~D. thesis, University of Wisconsin.
\bibitem[Shelton \& Cox 1994]{shelton95}Shelton, R.~L., \& Cox, D.~P. 1994, ApJ, {434}, 599.
\bibitem[Shull \& Van Steenberg 1982]{shull82}Shull, J.~M., \& Van Steenberg, M. 1982, ApJS., {48}, 95.
\bibitem[Shull \& Slavin 1994]{shull}Shull, J.~M., \& Slavin, J.~D. 1994, ApJ, {427}, 784.
\bibitem[Slavin \& Cox 1992]{slavin92}Slavin, J. \& Cox, D. 1992, ApJ, {392}, 131.
\bibitem[Slavin \& Cox 1993]{slavin93}Slavin, J. \& Cox, D. 1993, ApJ, {417}, 187.
\bibitem[Smith \& Cox 1974]{smith74}Smith, B.~W., \& Cox , D.~P. 1974, ApJ, {189}, L105.
\bibitem[Smith 1996]{smithr}Smith, R.~K. 1996, Ph.~D. Thesis, University of Wisconsin.
\bibitem[Spitzer 1956]{spitzer56}Spitzer, L. 1956, ApJ, {124}, 20.
\bibitem[Spitzer 1996]{spitzer96}Spitzer, L. 1996, ApJ, {458}, 29.
\bibitem[Welsh \etal 1989]{ssl}Welsh, B., Vallerga, J.~V., Jelinsky, P., Vedder, P.~W., Bowyer, S., \& Malina R.~F. 1989, Proc. SPIE, {1160}, 554.
\end{thebibliography}
\end{document}